\documentclass[twocolumn,english,showpacs]{revtex4}
\usepackage{babel,amsmath,amssymb,dcolumn}
\usepackage[dvips]{graphics}
\def\imo{i}
\begin{document}
\title{Gravitational spectrum of black holes in the Einstein-Aether theory.}
\author{R.A. Konoplya}
\email{konoplya@fma.if.usp.br}
\author{A. Zhidenko}
\email{zhidenko@fma.if.usp.br}
\affiliation{Instituto de F\'{\i}sica, Universidade de S\~{a}o Paulo \\
C.P. 66318, 05315-970, S\~{a}o Paulo-SP, Brazil}


\begin{abstract}
Evolution of gravitational perturbations, both in time and
frequency domains, is considered for a spherically symmetric black
hole in the non-reduced Einstein-Aether theory. It is shown that
real oscillation frequency and damping rate are larger for the
Einstein-Aether black hole than for the Schwarzschild black hole.
This may provide an opportunity to observe aether in the
forthcoming experiments with new generation of gravitational
antennas.
\end{abstract}

\maketitle

One of the most intriguing issues of modern physics consists in
attempts to go beyond local Lorentz symmetry \cite{LV}. In theory
of gravity, breaking of local Lorentz invariance leads to a
general relativity coupled to a dynamical time-like vector field
$u^{a}$, called ``aether''. More exactly, $u^{a}$ breaks local
boost invariance, while rotational symmetry in a preferred frame
is preserved \cite{AEreview}. Thereby, aether is a kind of locally
preferred state of rest at each point of space-time due-to some
unknown physics. Recently observable consequences of
Einstein-Aether theory attracted considerable interest
\cite{AEobserve}. Gravitational consequences of Local Lorentz
symmetry violation must show themselves in radiative processes
around black holes. It is known that gravitational radiation
damping of binary pulsars orbits reproduces the weak field general
relativity at lowest post- Newtonian order \cite{Foster}.  Yet,
the significant difference between Einstein and Einstein-Aether
theories should be seen in the regime of strong field, for
instance in observing  of the characteristic quasi-normal spectrum
of black holes. Thus, existence of aether could be tested in the
forthcoming experiments with new generation of gravitational
antennas. Motivated by the above reasons, in a previous letter
\cite{Konoplya-Zhidenko-PLB2} we developed a method for finding of
the quasinormal modes for the perturbations of metrics which are
not known analytically, but instead are given only numerically in
some region near black holes. That is the case of the
Einstein-Aether black holes found in \cite{Jacobson}. In
\cite{Konoplya-Zhidenko-PLB2}, there were found the quasinormal
modes for test scalar and electromagnetic fields in the vicinity
of the Einstein-Aether black holes. It was shown that the scalar
and electromagnetic quasinormal modes in the Einstein-Aether
theory, have larger real oscillation frequency and damping rate
than those of the Schwarzschild black holes in the Einstein
theory. As quasinormal spectrum does not depend on the spin of the
field in eikonal regime, qualitatively the same QN behavior was
suggested in \cite{Konoplya-Zhidenko-PLB2} for the gravitational
perturbations as for scalar and electromagnetic ones. In the
present work we show that it is indeed true, and analyze
gravitational perturbations of the Einstein-Aether black holes
both in frequency and time domain.

We shall start from the lagrangian of the full Einstein-Aether
theory forms the most general diffeomorphism invariant action of
the space-time metric $g_{ab}$ and the aether field $u^a$
involving no more than two derivatives given by
\begin{equation}\label{eaft}
L=-R-K^{ab}_{~~~mn}\nabla_au^m\nabla_bu^n-\lambda(g_{ab}u^au^b-1),
\end{equation}
here $R$ is the Ricci scalar, $\lambda$ is a Lagrange multiplier
which provides the unit time-like constraint,
$$K^{ab}_{~~~mn}=c_1g^{ab}g_{mn}+c_2\delta^a_m\delta^b_n+c_3\delta^a_n\delta^b_m+c_4u^au^bg_{mn},$$
where the $c_i$ are dimensionless constants.

Spherically symmetry allows to fix $c_4=0$. In this letter,
following
\cite{Jacobson}, we shall consider the so-called {\it non-reduced} Einstein-Aether
theory, for which $c_3=0$, and we can use the field redefinition
that fixes the coefficient $c_2$ \cite{Jacobson}:
$$c_2=-\frac{c_1^3}{2-4c_1+3c_1^2},$$
so that $c_1$ is the free parameter.

The metric for a spherically symmetric static black holes in
Eddington-Finkelstein coordinates can be written in the form
\cite{Jacobson}:
\begin{equation}
d s^{2} = N(r) d v^{2} - 2 B(r) d v d r - r^{2} d \Omega^{2},
\end{equation}
where the functions $N(r)$ and $B(r)$  are given by numerical
integration near the black hole event horizon \cite{Jacobson}. One
can re-write this metric in a Schwarzschild like form:
\begin{equation}\label{metric}
d s^{2} = - N(r) d t^{2} + \frac{B^{2}(r)}{N(r)}d r^{2} + r^{2} d
\Omega^{2}.
\end{equation}

Since the background value of aether coupling, determined by
constants $c_{i}$ is small in comparison with the background
characteristics of large black hole, determined by the mass of the
black hole $M$. Therefore, the background black hole metric is, in
fact, the Schwarzschild metric slightly corrected by the aether.
Thus, one can neglect small perturbations of aether, keeping only
linear perturbations of Ricci tensor. Then the perturbation
equations with unperturbed aether have the form:
\begin{equation}
\delta R_{\alpha \beta} =0.
\end{equation}
The general form of the perturbed metric, according to
Chandrasekhar designations, is
$$ d s^{2} = e^{2 \nu} d t^{2} - e^{2 \psi} (d \phi^{2} - \omega d t
-q_{2} d r - q_{3} d \theta)^{2}-$$
\begin{equation}
 e^{-2 \mu_{2}} d r^{2} - e^{-2
\mu_{3}} d \theta^{2}.
\end{equation}
Here
$$ e^{2 \nu} = N(r), \quad e^{2\mu_{2}} = N(r)/B^{2}(r), $$
\begin{equation}
e^{2\mu_{3}} =r^{2}, \quad e^{2 \psi} = r^{2} \sin^{2}\theta.
\end{equation}
Let us introduce new variables
\begin{equation}
Q_{ik} = q_{i, k} -q_{k, i}, \quad Q_{i0} = q_{i, 0} -\omega_{,i},
\quad i, k = 2, 3.
\end{equation}
Here we used $0, 1, 2, 3$ for $t$, $\phi$, $r$ and $\theta$
coordinates respectively. Here we shall consider the axial type of
gravitational perturbations. The equations (11),(12) on p.143 of
\cite{Chandra} can be reduced to a single equation
\begin{equation}
\frac{r^{4}}{B} \left(\frac{N}{B r^{2}} \frac{\partial Q}{\partial r}
\right)_{, r} -\frac{r^{2}}{B} \frac{\partial^{2} Q}{\partial
t^{2}} + \sin^{3} \theta \frac{\partial }{\partial \theta}
\left(\frac{1}{sin^{3} \theta} \frac{\partial Q}{\partial \theta}
\right)=0
\end{equation}
where we used
\begin{equation}
Q =e^{3 \psi + \nu - \mu_{2} -\mu_{3}} Q_{23}.
\end{equation}
The following representation of the function $Q(t, r, \theta)$
\begin{equation}
Q(t, r, \theta) = r R_{\ell}(t, r) C_{\ell + 2}^{-3/2}(\theta)
\end{equation}
leads to separation of angular variable $\theta$.

Finally we have the wave-like equation for the radial coordinate
\begin{equation}\label{wavelike}
\frac{d^{2} \Psi}{d r_{*}^{2}} + (\omega^{2} - V(r))\Psi =
0,\qquad d r_{*}=\frac{B(r)}{N(r)}dr.
\end{equation}
with the effective potential
\begin{equation}\label{potential}
V(r) = N(r) \frac{(\ell + 2)(\ell -1)}{r^{2}} + \frac{2
N^{2}(r)}{B^{2}(r) r^{2}} - \frac{1}{r} \frac{d
(N(r)/B(r))}{dr^{*}}.
\end{equation}

Quasi-normal modes of asymptotically AdS black holes have been
studies recent years extensively, because of their interpretation
in Conformal Field Theory \cite{CFT} with some specific boundary
conditions. In astrophysically relevant problem, one should
require natural boundary conditions for QN modes of purely
in-going waves at the event horizon and purely out-going waves at
spatial infinity
\begin{equation}
\Psi \sim  e^{\pm i r_{*} \omega}\quad r_{*} \rightarrow \pm \infty.
\end{equation}
Under these boundary conditions, the quasinormal modes were
studied in a great number of papers \cite{QNMs}, yet in those
cases the background metric and the effective potential were known
in analytical form. For the case of Einstein-Aether theory, we are
in position to apply the method developed in our previous paper
\cite{Konoplya-Zhidenko-PLB2}. Here we shall give only a brief
summary of the whole procedure of \cite{Konoplya-Zhidenko-PLB2}.

We approximate the numerical data for the metric by a fit of the
form
\begin{equation}\nonumber
N(r) =
\frac{\displaystyle\sum_{i=0}^{N_N}a^{(N)}_ir^{i}}{\displaystyle1 +
  \sum_{i=1}^{N_N}b^{(N)}_i r^{i}}, \quad B(r) =
\frac{\displaystyle\sum_{i=0}^{N_B}a^{(B)}_ir^{i}}{\displaystyle1 +
  \sum_{i=1}^{N_B}b^{(B)}_i r^{i}}.
\end{equation}
which are substituted into equations (\ref{wavelike}) and
(\ref{potential}). The numbers $N_N$ and $N_B$ determine the
number of terms in the polynomials and are chosen in order to
provide best convergence of the WKB series. Coefficients
$a^{(N)}_i$, $b^{(N)}_i$, $a^{(B)}_i$, $b^{(B)}_i$ are determined
by the fitting procedure. The WKB expansion has the form
\begin{equation}\label{WKBformula}
\frac{\imath Q_{0}}{\sqrt{2 Q_{0}''}}
- \sum_{i=2}^{6} \Lambda_{i} = n+\frac{1}{2},
\end{equation}
where the correction terms of the i-th WKB order $ \Lambda_{i}$
can be found in \cite{WKB,Will-Schutz} and
\cite{Konoplya-prd3}, $Q = V - \omega^2$ and $Q_{0}^{i}$ means the
i-th derivative of $Q$ at its maximum.

Alternatively, we shall use the above mentioned fits of the metric
functions, and consequently of the effective potential, in the
time-domain analysis: using the integration scheme described for
instance in \cite{Price-Pullin}. In detail, we used a numerical
characteristic integration scheme, based in the light-cone
variables $u = t - r_\star$ and $v = t + r_\star$. In the
characteristic initial value problem, initial data are specified
on the two null surfaces $u = u_{0}$ and $v = v_{0}$. The
discretization scheme applied, is
\begin{eqnarray}
\lefteqn{\Psi(N) = \Psi(W) + \Psi(E) -
\Psi(S) }  \nonumber \\
& & \mbox{} - \Delta^2 V(S) \frac{ \Psi(W) + \Psi(E)}{8} +
\mathcal{O}(\Delta^4)   \ ,
\label{d-uv-eq}
\end{eqnarray}
where we have used the definitions for the points: $N = (u +
\Delta, v + \Delta)$, $W = (u + \Delta, v)$, $E = (u, v + \Delta)$
and $S = (u,v)$.

The application of the above two methods shows excellent
agreement: for instance the fundamental mode $0.7686-0.1887
\imo$ in time domain is very close to the WKB value
$0.769470-0.187783\imo$ for $c_{1} =0.4$, as can be seen in Fig.
2. From the obtained numerical data in Table I and time domain
pictures in Fig. 1-2, one can see that when increasing $c_{1}$,
both real oscillation frequency and damping rate are increasing.
Even for a small aether $c_{1} \sim 0.1$, the increase in $Re
\omega$ and $Im \omega$ is of about half percent, and could, in
principle, be detected by new generation of gravitational
antennas. For larger $c_{1}$, the difference between,
Schwarzschild and Einstein-Aether QNMs can be very significant and
reach six-seven percents.  From the Table I, it is evident that
both real and imaginary parts of $\omega$ grows when increasing
the multipole number $\ell$. Here we considered only axial
gravitational perturbations, which are iso-spectral with polar
gravitational perturbations for Schwarzschild black holes. For
black holes in Einstein-Aether theory this iso-spectrality will be
broken, and QNMs for polar perturbations should slightly differ
from axial, when considering the full perturbations of
Einstein-Aether equations. The same breaking of iso-spectrality
happens, for instance, when perturbing dilaton black holes or
black holes in higher than four dimensional space-times
\cite{Konoplya:2003dd}. In our approach, the perturbations of
aether were neglected in comparison with perturbations of the
metric of a large astrophysical black hole. Therefore this
difference between axial and polar QN spectra was neglected as
well.

Note that we used here the method based on the supposition that QN
frequencies are determined mainly near the peak of the potential
barrier, while behavior of the potential barrier far from black
hole is not significant. Even despite this idea was inspired by
WKB approach, it is not dependent on WKB technique, as was shown
here by computations in time domain.

\begin{widetext}

\begin{table}
\caption{Axial gravitational perturbations for the non-reduced Einstein-Aether theory: The fundamental mode.}
\begin{tabular}{|l|c|c|c|c|}
\hline
$c_1$&$\ell=2$&$\ell=3$&$\ell=4$&$\ell=5$\\
\hline
$0.1 $&$0.751958-0.179578\imo$&$1.206905-0.187387\imo$&$1.629395-0.190378\imo$&$2.038526-0.191824\imo$\\
$0.2 $&$0.757144-0.180786\imo$&$1.215669-0.188850\imo$&$1.641492-0.191932\imo$&$2.053816-0.193422\imo$\\
$0.3 $&$0.762871-0.184443\imo$&$1.225598-0.192601\imo$&$1.655131-0.195746\imo$&$2.071021-0.197268\imo$\\
$0.4 $&$0.769470-0.187783\imo$&$1.236966-0.196172\imo$&$1.670792-0.199412\imo$&$2.090803-0.200981\imo$\\
$0.5 $&$0.777059-0.192176\imo$&$1.250474-0.200794\imo$&$1.689421-0.204151\imo$&$2.114345-0.205777\imo$\\
$0.6 $&$0.786784-0.198087\imo$&$1.267215-0.207061\imo$&$1.712545-0.210574\imo$&$2.143595-0.212276\imo$\\
$0.7 $&$0.799391-0.206961\imo$&$1.289844-0.216264\imo$&$1.743878-0.220000\imo$&$2.183271-0.221813\imo$\\
$0.77$&$0.811083-0.216302\imo$&$1.311449-0.225807\imo$&$1.773893-0.229784\imo$&$2.221342-0.231720\imo$\\
\hline
\end{tabular}
\end{table}

\end{widetext}

\begin{figure}\label{aether.time1}
\caption{Evolution of axial gravitational perturbations in time domain $c_{1}=0.4$, $\ell =3$, non-reduced
theory (red line) in comparison with the Schwarzschild case (blue
line).}
\resizebox{\linewidth}{!}{\includegraphics*{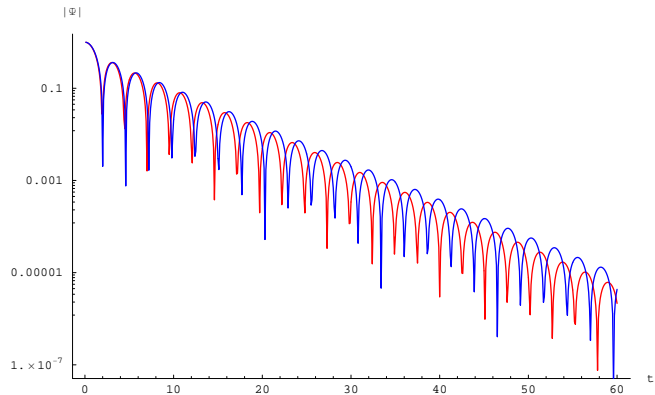}}
\end{figure}

\begin{figure}\label{aether.time2}
\caption{Evolution of axial gravitational perturbations in time domain $c_{1}=0.1$ (green line), $c_{1}=0.4$
(red line), $\ell =2$, non-reduced theory in comparison with the
Schwarzschild case (blue line).}
\resizebox{\linewidth}{!}{\includegraphics*{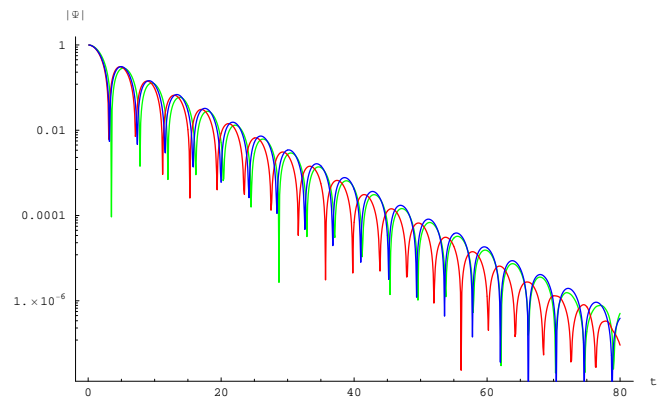}}
\end{figure}

\begin{acknowledgments}
This work was supported by \emph{Funda\c{c}\~{a}o de Amparo
\`{a} Pesquisa do Estado de S\~{a}o Paulo (FAPESP)}, Brazil.
\end{acknowledgments}

\end{document}